# 1D self-healing beams in integrated silicon photonics

Zhuoran Fang[1*], Rui Chen[1], Albert Ryou[1], Arka Majumdar[1,2,*]

[1]Department of Electrical and Computer Engineering, University of Washington, Seattle, WA 98195, USA
[2]Department of Physics, University of Washington, Seattle, WA 98195, USA
*rogefzr@uw.edu, arka@uw.edu

**Abstracts**

Since the first experimental observation of optical Airy beams, various applications ranging from particle and cell micromanipulation to laser micromachining have exploited their non-diffracting and accelerating properties. The later discovery that Airy beams can self-heal after being blocked by an obstacle further proved their robustness to propagate under scattering and disordered environment. Here, we report the generation of Airy beams on an integrated silicon photonic chip and demonstrate that the on-chip 1D Airy beams preserve the same properties as the 2D beams. The 1D meta-optics used to create the Airy beam has the size of only $3\mu m \times 16\mu m$, at least three orders of magnitude smaller than the conventional optic. The on-chip self-healing beams demonstrated here could potentially enable diffraction-free light routing for on-chip optical networks and high-precision micromanipulation of bio-molecules on an integrated photonic chip.

**Introduction**

Non-diffracting beams, exemplified by the well-known Bessel beam[1], have long been pursued in the photonics community to avoid the ubiquitous phenomenon of diffraction in Gaussian beams. In practice, these "non-spreading" beams are normally truncated by an aperture, so they do tend to diffract during propagation[2]. However, by using an appropriate size of the aperture, the diffraction can be significantly slowed down over the intended propagation distance and for many practical applications they can be considered "diffraction-free". Unlike the straight propagating Bessel beams, Airy beams emerge as an alternative non-diffracting solution to the paraxial wave equation that exhibits parabolic trajectory during propagation[3]. Airy beams were first predicted and experimentally observed in the optical domain by Siviloglou and Christodoulides[2,4] and subsequently demonstrated in electrons[5] and plasmons[6]. It was soon discovered that Airy beams can reconstruct themselves after being blocked by an obstacle or propagating through a turbulent environment[7]. Following these experimental observations, numerous works have exploited the non-diffracting, self-bending, and self-healing properties of Airy beams for applications in microcopy[8,9], particle and cell manipulation[10–12], optical trapping[12], micro and nano machining[13,14] and optical filamentation[15,16].

Traditionally, Airy beams are generated by reflecting light from a cubic phase mask, usually realized by a spatial light modulator, and then focused by a spherical lens in free space. This generally involves bulky optics and requires precise alignment, which preclude the miniaturization of the whole optical systems. In this paper, we demonstrate for the first time the generation of self-healing Airy beams in integrated silicon photonics near the telecommunication wavelength using an on-chip meta-optic. We show that the 1D on-chip Airy beams preserve the non-diffracting and self-healing properties of 2D Airy beams. The on-chip meta-optic used to generate the 1D Airy beams has the size of $3\mu m \times 16\mu m$ which is almost three orders of magnitude smaller than the conventional macroscopic optics. Furthermore, no alignment of optical components is required as the meta-optics combine the functionality of a cubic phase-mask and the lens in a single device. The integrated meta-optics can provide alignment with other optical components on-chip with nano-meter scale precision controlled by the lithography. Finally, we showed that the on-chip Airy beam can self-heal, which can provide a way to efficiently route light in the presence of fabrication imperfections and will be beneficial for large-scale on-chip optical interconnects. Future applications of our on-chip Airy beams include $1 \times N$ switches for on-chip light routing[17], high precision optical trapping[12], and particle manipulation[10] on a chip.

**Results**

Drawing on the design principle proposed by Wang[18] and Zhan[19], we incorporate the cubic phase mask into the quadratic lens phase mask to generate 1D Airy beams using a 1D phase function:

$$\phi(y) = \frac{2\pi}{\lambda} n_{eff} \left( f - \sqrt{f^2 + y^2} \right) + \alpha y^3 \quad (1)$$

where $\lambda$ is the operating wavelength, $n_{eff}$ is the effective index of silicon-on-insulator (SOI) waveguide, $f$ is the focal length of the 1D metalens, $y$ is the lateral extent of the meta-optics, and $\alpha$ controls the strength of the cubic term relative to the quadratic term. Such a phase profile can be readily implemented by tuning the length of the air slot etched into the silicon slab (Fig. 1a) on silicon dioxide. A focal length of 15µm is chosen to provide a good trade-off between the focusing efficiency and the collected light intensity. The range over which the Airy beam profile appears, also known as the depth of focus, can be controlled by adjusting α: increasing α increases the depth of focus as the cubic phase becomes more dominant. However, a larger α will also lead to broader main lobe and increased side lobes (Supplementary information Fig. S1) which compromises the efficiency of the meta-optics since a focused main lobe is generally desired for most practical applications[10–12]. For example, at α = 0.025 the Airy beam stays focused from $\Delta x$=10~20µm but becomes broad and exhibits multiple side lobes across the same range at α = 0.1(Supplementary information Fig. S1), $\Delta x$ being the distance between the input and output port as shown in Fig.1b. Here, α is chosen to be 0.025 to ensure that the depth of focus includes the focal point ($\Delta x$=15µm) to ensure that the main lobe intensity of the Airy beam is large, since the underlying Gaussian beam is also focused in the same range. This is particularly important for experimental measurements, as stronger light intensity ensures higher signal to noise ratio. Fig. 1c compares the simulated 2D intensity profile of the light propagating through the quadratic (top) and cubic (bottom) 1D meta-optics at 1550nm. The pitch Λ as indicated in Fig. 1a is chosen to be 500nm and the duty cycle η (defined as $\eta = \frac{w}{\Lambda}$, where $w$ is the width of the air slot) is 0.215 for both meta-optics, ensuring high transmission through the air slot[18]. For this combination of Λ and η, $n_{eff}$ is calculated to be 2.18 using finite difference time domain (FDTD) simulation. The quadratic metalens implements only the quadratic part of the phase mask defined by equation (1) while the cubic meta-optics implements the whole phase. Light is focused by the quadratic metalens at around Δ$x$=15μm but quickly diverges. On the other hand, the cubic meta-optics produces a curved beam with an extended depth of focus (EDOF), a defining signature of an Airy beam (Fig. 1c).

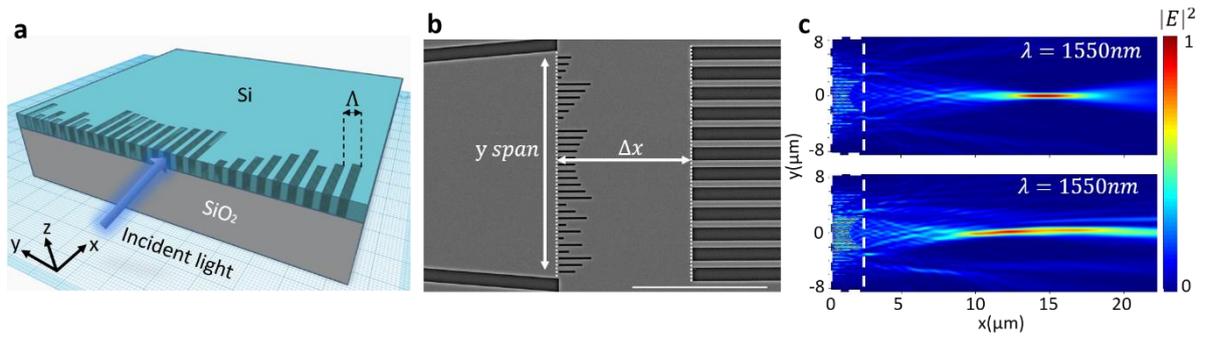

**Figure 1: Design of on-chip cubic meta-optics.**

**a** Schematic showing the structure of 1D cubic meta-optics. The blue arrow indicates the direction in which the light is launched into the metalens. **b** SEM image of a cubic metalens on SOI (scale bar:10μm). **c** Simulated electric field intensity of light focused by a quadratic meta-optics (top) and cubic meta-optics (bottom). The focal length is 15μm for both metasurfaces and $\alpha$ of the cubic meta-optics is 0.025. The white dashed boxes highlight where the phase masks are.

The simulation is experimentally verified using an integrated silicon photonic chip (see Materials and methods for fabrication details) and the results are shown in Fig. 2. Figs. 2a and 2b compare the 1D intensity distribution of light propagating different distances of Δ$x$ from a quadratic metalens and a cubic meta-optic, respectively. The optical transmission is collected across the $y$ span of the meta-optics using a vertical fiber setup via multiple waveguides

(indicated in Fig. 1b; see Materials and methods). A typical Gaussian beam profile can be seen from Fig. 2a, matching the simulation. The beam focuses tightly near $\Delta x$=15μm (FWHM~0.7 μm), begins to diverge at $\Delta x$=20μm, and diverges remarkably at $\Delta x$=30μm (FWHM~6 μm). In contrast, the beam originating from the cubic meta-optics maintains a narrow FWHM of ~1 μm across the same propagation distance (Fig. 2b), particularly at $\Delta x$=30μm where the Gaussian beam diverges significantly and becomes very broad. The asymmetric beam profile and the formation of side lobes in Fig. 2b are characteristics of Airy beams. Furthermore, an interesting phenomenon is observed at $\Delta x$=30μm, where a secondary lobe starts to form next to the main lobe. This is a well-known feature of Airy beam where the main lobe will split into secondary lobes as the beam evolves[7,10]. The secondary lobe will become the new main lobe while the previous main lobe will evolve into the side lobes. This property is the fundamental reason behind self-healing of Airy beam, which will be discussed later. Although the 1D Airy beam is not as tightly focused as the Gaussian beam at the focal length, it exhibits a moderately intense beam across a wider range. This could be beneficial for beam routing applications where different wavelengths of light or objects at different locations along the optical axis can be equally focused by the meta-optics, as exploited for imaging via 2D Airy beam[20]. Note that the experimental results at $\Delta x$=30μm for quadratic meta-optics (Fig. 2a) deviate more from the simulation than the results for Airy beam (Fig. 2b). We primarily attribute this to the significant increase of side lobes from the stronger diffraction of Gaussian beam. The limited resolution of output ports determined by the waveguide spacings (1.5 μm) fails to capture the multiple diffraction lobes accurately. Additionally, the reduction of signal to noise ratio could also contribute to the inaccuracy. In fact, we found in experiments that the light intensity of Gaussian main lobe drops to approximately 60% of that of Airy main lobe at $\Delta x$=30μm. Nevertheless, we clearly observe the non-diffracting nature of 1D Airy beam.

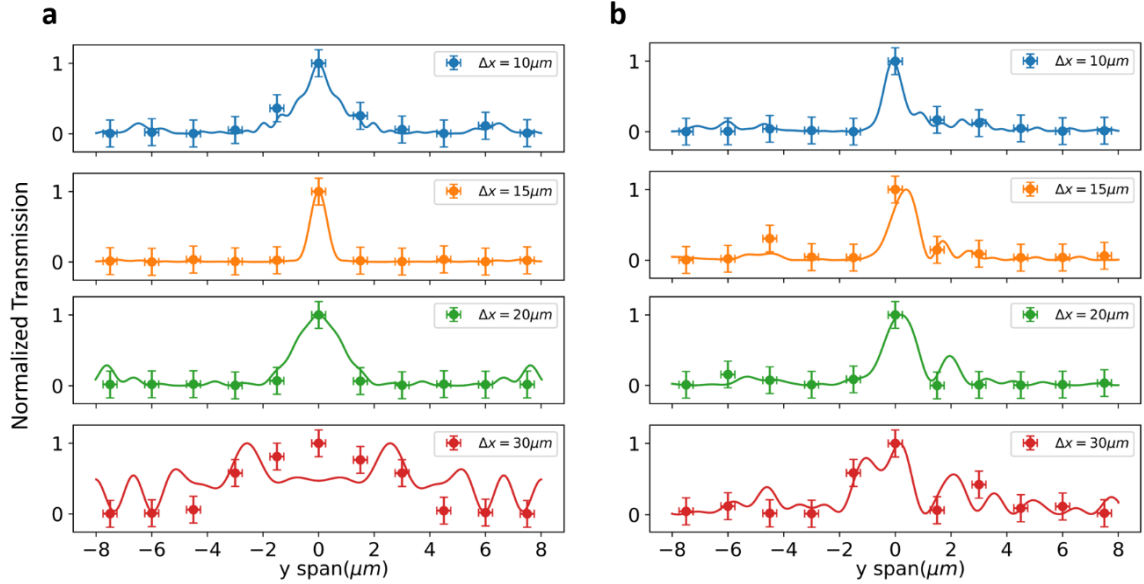

**Figure 2: Experimental results of 1D non-diffracting Airy beam.**

**a** Intensity distribution across the y-span of light focused by a quadratic metalens. Optical transmission is measured at $\Delta x$=10μm, 15μm, 20μm and 30μm. **b** Intensity distribution across the y span of light focused by a cubic meta-optic. Optical transmission is measured at $\Delta x$=10μm, 15μm, 20μm and 30μm. Solid lines are the simulation results and dots are the measured data in experiment at each waveguide. Error bars in *x* are determined by the lateral span of the intensity point which equals to the 500nm waveguide width. Error bars in *y* are calculated from the standard deviation of grating coupler efficiency across the whole chip.

We then experimentally verify the self-healing properties of Airy beam (see Fig. 3). Here, a circular air pocket (centered at $\Delta x$=8.8μm, *y*=0 μm) of 1μm radius is used to scatter the main lobes of the Gaussian beam and Airy beam, as shown in Fig. 3a and Fig. 3d, respectively. Optical transmission is measured at $\Delta x$=10μm, 20μm, and 30μm where the position and size of the scatterer is fixed. The simulated 2D electric field profile is superimposed on the SEMs of the meta-optics along with the scatterers. Strong scattering occurs as the Gaussian beam interacts with the air pocket, forming multiple symmetric side lobes (Fig. 3a). The beam scatters around the air cylinder and focuses again but the intensity deteriorates dramatically compared to Fig. 1c. At $\Delta x$=30μm, the main lobe of the Gaussian beam becomes almost indistinguishable as most energy scatters into the side lobes. This phenomenon is also observed experimentally for both air and Aluminum scatterer shown in Fig. 3b and Fig. 3c. The Gaussian main peak is completely suppressed at $\Delta x$=10μm and partially reforms with multiple side peaks at $\Delta x$=20μm. At $\Delta x$=30μm, the energy in the main lobe is completely transferred to the low-intensity side lobes due to the strong scattering. Replacing the air scatterer with Aluminum yields similar results (Fig. 3c). On the other hand, cubic meta-optic produces beams that are scattered less prominently by the same air pocket (Fig. 3d), because the intensity of the Airy beam is distributed more uniformly along the transverse direction. It can be clearly seen from the electric filed profile in Fig. 3d that the side lobes not blocked by the air pocket reform the main lobe after propagation over a certain distance, as indicated by the white arrow. This phenomenon matches the underlying principle of self-healing of Airy beams[10,21]. Distinct rays contribute to the main lobes at different propagation distances. When the main lobe is blocked, the unblocked rays from the side lobes will reconstitute the main lobe after a "healing distance"[7]. The experimental data in Fig. 3e shows that at $\Delta x$=10μm, the main lobe is completely blocked whereas the side lobes are not. The side lobes become more intense and merge to form the main lobe at $\Delta x$=20μm as the self-healing gradually takes place. At $\Delta x$=30μm, the main lobe reforms completely when the maxima re-emerges at the center $y$=0μm. Fig. 3f shows the experimental results of 1D Airy beam being blocked by an Al scatterer. Since almost no light from the main lobe is able to penetrate the metal, the side lobes intensity becomes weaker at $\Delta x$=20μm compared to the air scatterer at the same traveling distance. As the beam gradually propagates and reconstructs itself, the intensity of the main lobe recovers, as can be seen at $\Delta x$=30μm in Fig. 3f. In fact, among the three different propagation lengths, the main lobe at $\Delta x$=30μm carries the most power. The recovery of the beam shape and power in the main lobe is sign of complete self-healing of the Airy beam. On the contrary, the Gaussian beam has completely decayed and lost its shape at the same propagation distance (Fig. 3c).

In Table 1, we calculate the absolute values of Pearson correlation coefficients[22] between the scattered and unscattered Gaussian/Airy beams to quantitively analyze the self-healing behavior. Pearson correlation coefficients measure the linear correlation between two sets of data X and Y and have values from -1 to 1. Zero indicates no linear correlation at all while 1 implies that a linear equation describes the relationship between X and Y perfectly. Hence, the closer the coefficient's absolute value to one, the stronger the association between the two data sets. Here, X is the unscattered beam profile shown in Fig. 2 and Y is the scattered beam profile shown in Fig. 3. At $\Delta x$=10μm, both Gaussian and Airy beams in presence of scatterer correlate poorly with the unscattered beams, indicating a large disparity between their profiles. At $\Delta x$=20μm, the scattered Gaussian beam matches the unscattered Gaussian beam profile almost perfectly with a correlation coefficient as high as 0.97. This is due to the insufficient blocking of the light as shown in the simulation (Fig. 3a). On the other hand, the correlation of scattered Airy beam remains to be near zero as self-healing is incomplete. At $\Delta x$=30μm, since there is no self-healing, the scattered Gaussian beam deviates significantly from the unscattered beam as its correlation coefficient drops to 0.5 for air pocket and 0.35 for Al scatterer. In contrast, thanks to the self-reconstruction of Airy beam, the correlation coefficient of Airy beam increases dramatically to 0.84 for air pocket and 0.77 for Al scatterer as the scattered beam almost resembles the unscattered beam.

Table 1. Calculated absolute values of Pearson correlation coefficients between the experimentally measured unscattered and scattered beam profiles

| $\Delta x$(μm) | Gaussian, air | Gaussian, Al | Airy, air | Airy, Al |
| --- | --- | --- | --- | --- |
| 10 | 0.21 | 0.22 | 0.02 | 0.00 |
| 20 | 0.97 | 0.96 | 0.08 | 0.01 |
| 30 | 0.50 | 0.35 | 0.84 | 0.77 |

Finally, it should be pointed out that the experimental data deviates more significantly from the simulation when the Al scatterer is used. This can be attributed to the non-ideal packing of thermally evaporated Al inside the air pocket. From the SEM characterization, we found that the lift-off of Al results in a truncated cone, instead of a perfect cylinder, due to the shadowing effect of evaporation. Hence, some light is still able to penetrate the scatterer, producing a mixed scattering pattern. Despite this, the Al scatterer sufficiently attenuates the light more than air to allow the revelation of self-healing.

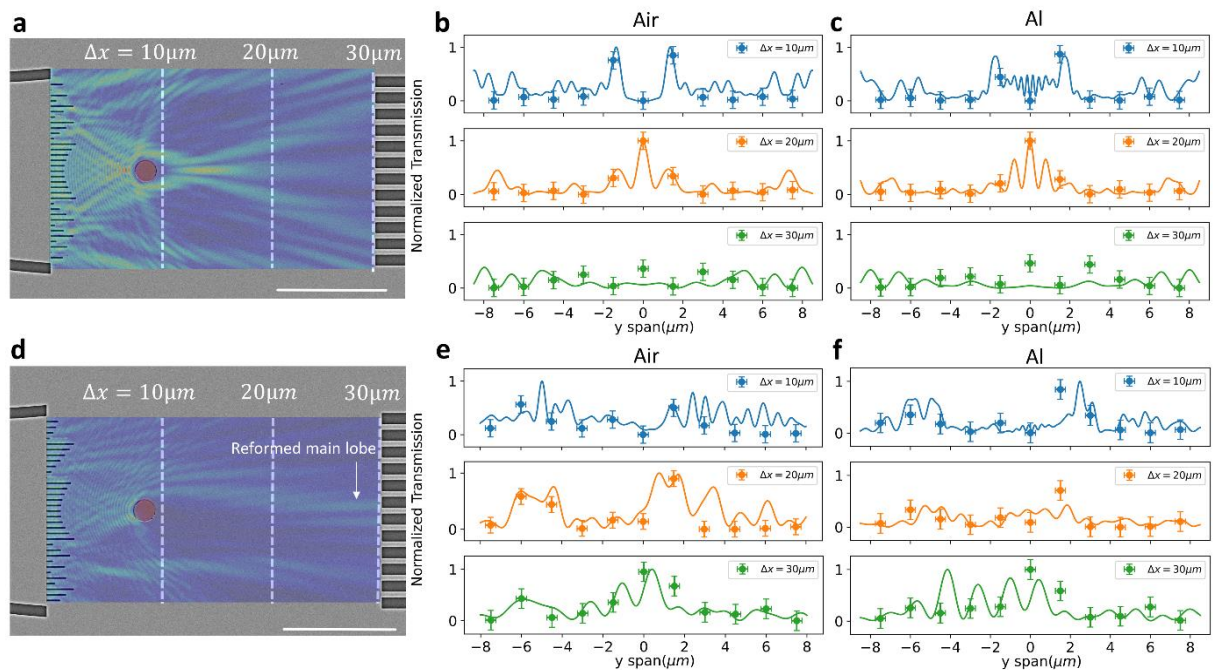

**Figure 3: Self-healing of on-chip 1D Airy beam.**

**a** SEM showing the quadratic metalens and the circular air pocket (scale bar: 10μm). The scatterer is highlighted with false color and the simulated electric field distribution is overlapped onto the SEM image. The white dashed lines indicate where the optical transmission is measured. **b** Intensity distribution across the y span showing the scattering of Gaussian beam by a circular air hole. **c** Intensity distribution across the y span showing the scattering of Gaussian beam by a circular Al scatterer. **d** SEM showing the cubic meta-optic and the circular air pocket (scale bar: 10μm). The reformed Airy main lobe is indicated by an arrow. **e** Intensity distribution across the y span showing the scattering of Airy beam by a circular air pocket. **f** Intensity distribution across the y span showing the scattering of Airy beam by a circular Al scatterer. Solid lines are the simulation results and dots are the measured data in experiment at each waveguide. Error bars in $x$ are determined by the lateral span of the intensity point which equals to the 500nm waveguide width. Error bars in $y$ are calculated from the standard deviation of grating coupler efficiency across the whole chip.

We further explore the self-healing of the side lobes in an Airy beam. A rectangular air slot ($1\mu m \times 2\mu m$, centered at $\Delta x=12.2\mu m$, $y=2\mu m$) is used to block the side lobes of the 1D Airy beam (Fig. 4a). It can be clearly seen that the side lobes disappear in the shadow region behind the rectangular slot but starts to recover at $\Delta x=15.5\mu m$. At $\Delta x=20.5\mu m$, multiple side lobes have recovered their original intensity. Similar to the self-healing of the main lobe, the side lobes can self-heal, because the rays from the unblocked main lobe evolve into side lobes as they propagate. The simulation is corroborated by the experiments (Fig. 4b). Note that the main peak of the Airy beam is unaffected as the scatterer is placed off-center. Near the air slot, the side lobes are completely suppressed (top plot, Fig. 4b). From $\Delta x=15.5\mu m$ to $20.5\mu m$, the primary side peak near $y=1.5\mu m$ gradually reforms, matching the prediction of the simulation.

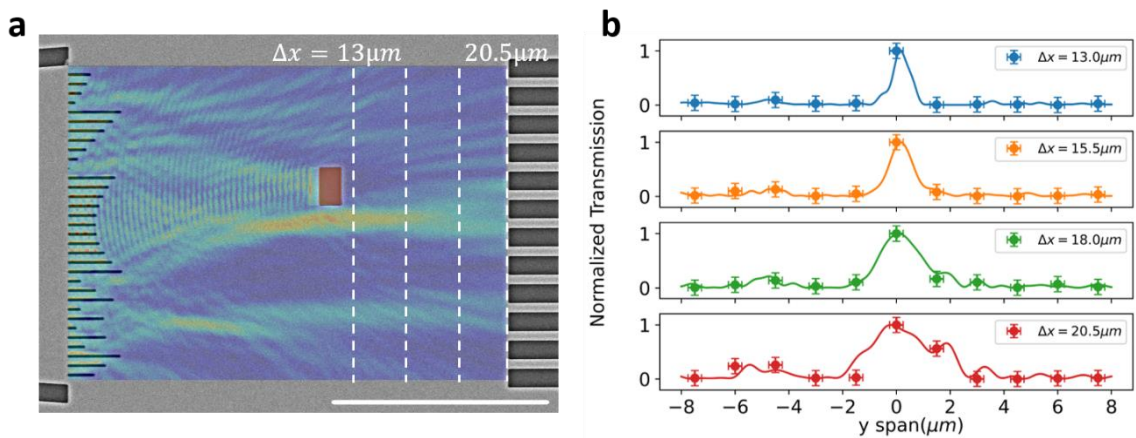

**Figure 4: Self-healing of on-chip 1D Airy beam side lobes**

**a** SEM of the cubic meta-optic and the rectangular air scatterer (scale bar: 10µm). The scatterer is highlighted with false color and the simulated electric field distribution is overlapped onto the SEM image. The white dashed lines indicate where the optical transmission is measured. **b** Intensity distribution across the y span showing the self-healing of Airy beam side peaks. Solid lines are the simulation results and dots are the measured data in experiment measured at each waveguide. Error bars in *x* are determined by the lateral span of the intensity point which equals to the 500nm waveguide width. Error bars in *y* are calculated from the standard deviation of grating coupler efficiency across the whole chip.

Finally, we show broadband operation across the telecommunication C-band with the cubic meta-optics. To understand the origin of the broadband behavior, it is essential to first analyze how the depth of focus and focal length of a 1D metalens is influenced by the chromatic dispersion. To begin with, the depth of focus of an ordinary lens with width $w$ and focal length $f$ is given by[23]:

$$\Delta f = 4\lambda \frac{f^2}{w^2}$$

For a 1D metalens of 16μm width and focal length of 15μm operating at 1550nm, the depth of focus is calculated to be ~5.45μm. Such depth of focus can be further increased by designing a smaller diameter lens with long focal length. Meanwhile, the focal length shift of a 1D metalens from wavelength variation originates from the fixed $\phi(\lambda) = \frac{2\pi}{\lambda}L$, imposed by length $L$ of the air slot, if we assume the silicon is approximately dispersionless in the wavelength of interest. At a different wavelength $\lambda_1$, the phase imparts on the wavefront becomes $\phi(\lambda_1)$ instead of the designed phase $\phi(\lambda_0)$. At the locations where phase wrapping occurs, $L$ changes abruptly and hence $\phi(\lambda_1)$ can no longer follow the ideal phase profile for $\lambda_1$ but approaches the phase profile $\phi(\lambda_0)$[24]. As a result, the focal length changes with the wavelength shift with an approximate relationship of $f \propto \frac{1}{\lambda}$. We estimate that the focal length changes from 15.28μm at 1525nm to 15.08μm at 1545nm. This focal length change of 0.2μm is far less than the calculated depth of focus of 5.45μm, which means the Gaussian beam remains tightly focused despite the wavelength shift and the metalens is fundamentally broadband in the telecom C-band. This is experimentally verified, accompanied with FDTD simulation in Figure 5c (measured at $\Delta x$=15μm). Such broadband behavior has also been reported in simulation and experiment before[18,25].

Fig. 1 and Fig. 2 have already shown that the addition of cubic phase can extend the depth of focus of a metalens, which makes the focal length drift due to dispersion even less noticeable. This is confirmed in Fig. 5b. Although the cubic meta-optic is designed to operate at 1550nm, varying the center wavelength does not affect its performance. The 1D Airy beam profile

(measured at $\Delta x=15\mu m$) originated from the same cubic metalens remains relatively unchanged across the telecom C-band and the main lobe does not diverge significantly thanks to its EDOF. We note that, however, the EDOF further increases the bandwidth of the cubic meta-optics and ultimately the operation bandwidth will only be limited by the grating coupler efficiencies.

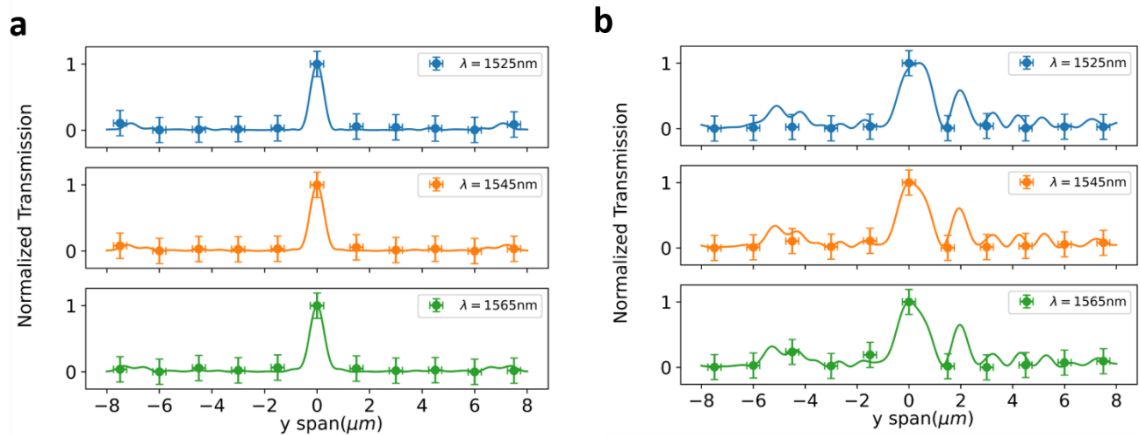

**Figure 5: Broadband operation of cubic meta-optics across the C-band**

**a** Gaussian beam profile formed by a quadratic metalens measured at $\Delta x=15\mu m$ across the telecommunication C-band. **b** 1D Airy beam profile measured at $\Delta x=15\mu m$ across the telecommunication C-band. Solid lines are the simulation results and dots are the measured data in experiment at each waveguide. Error bars in $x$ are determined by the lateral span of the intensity point which equals to the 500nm waveguide width. Error bars in $y$ are calculated from the standard deviation of grating coupler efficiency across the whole chip.

**Discussion:**

In conclusion, we have demonstrated for the first time the formation of self-healing and non-diffracting Airy beam in a waveguiding structure using silicon integrated photonics. It should be pointed out that such 1D Airy beam is not dependent on the material platform. Any planar waveguides can be designed to support the propagation of 1D Airy beam across different wavelengths of interest. For example, recently a broadband 1D metalens has been designed on $Al_2O_3$ platform for operation in the visible wavelength[25]. The non-diffracting and self-healing nature of 1D Airy beam makes it particularly suitable for operation in turbulent and scattering media such as fluid. Hence, the 1D Airy beam could possibly be generated in optofluidic waveguides[26] for applications such as microfluidic flow cytometry[17,27] and cell sorting[28]. Additionally, the methodology we present here is universal and hence other classes of non-diffracting beams such as Bessel beam[29] can also be generated by tailoring the phase mask design. Finally, reconfigurability can be explored in the future using integrated metallic heaters[17] or phase change materials[30,31] for beam steering and focal length tuning of the 1D self-healing beam, which can act as an ultra-compact $1 \times N$ optical switch for on-chip diffraction-free light routing.

**Materials and methods:**

*SOI Device Fabrication:* The 1D cubic and quadratic lenses were fabricated on a 220-nm thick silicon layer on top of a 3-μm-thick buried oxide layer (SOITECH). The pattern was defined by a JEOLJBX-6300FS 100kV electron-beam lithography (EBL) system using positive tone ZEP-520A resist. 220 nm fully etched ridge waveguides, meta-optics, and air scatterers were made by an inductively coupled plasma reactive ion etching (ICP-RIE) process. A second EBL exposure using positive tone poly(methyl methacrylate) (PMMA) resist was subsequently carried out to create windows for the Al deposition. After development, 220nm Al was electron beam evaporated onto the chip. The lift-off of Al was completed using methylene chloride.

*Experimental setup:* The meta-optics on SOI were characterized by a vertical fiber-coupling setup[30]. All the measurements were performed under ambient conditions while the temperature of the stage was fixed at ~24°C by a thermoelectric controller (TEC, TE Technology TC-720). The input light was provided by a tunable continuous-wave laser (Santec TSL-510) and its polarization was controlled by a manual fiber polarization controller (Thorlabs FPC526) to match the fundamental quasi-TE mode of the waveguides. A low-noise power meter (Keysight

81634B) was used to collect the static optical output from the grating couplers.

**Conflict of interest statement**

The authors declare no conflict of interest.

**Acknowledgements**

Z.F. and A.M. conceived the project. Z.F. simulated, designed, and fabricated the devices. Z. F. performed the experiments. R.C. helped with the experiments and simulations. A.R. helped with the simulation. A.M. supervised the overall progress of the project. Z.F. wrote the manuscript with input from all the authors.

The research is funded by National Science Foundation (NSF-2003509), ONR-YIP Award, and Washington Research Foundation. Part of this work was conducted at the Washington Nanofabrication Facility / Molecular Analysis Facility, a National Nanotechnology Coordinated Infrastructure (NNCI) site at the University of Washington, which is supported in part by funds from the National Science Foundation (awards NNCI-1542101, 1337840 and 0335765), the National Institutes of Health, the Molecular Engineering & Sciences Institute, the Clean Energy Institute, the Washington Research Foundation, the M. J. Murdock Charitable Trust, Altatech, ClassOne Technology, GCE Market, Google, and SPTS.